\documentclass[aps,pra,twocolumn,groupedaddress,showpacs]{revtex4}
\pdfoutput=1
\usepackage{subfigure,epsfig,amsfonts,amsmath,graphicx,dcolumn,bm}

\begin{document}

\title{Photon Routing in Cavity QED: Beyond the Fundamental Limit of Photon Blockade}

\author{Serge Rosenblum$^1$}
\author{Scott Parkins$^2$}
\author{Barak Dayan$^1$}
\affiliation{$^1$Department of Chemical Physics, Weizmann Institute of Science, Rehovot 76100, Israel}
\affiliation{$^2$Department of Physics, University of Auckland, Auckland, New Zealand}

\date{\today}

\begin{abstract}
The most simple and seemingly straightforward application of the photon blockade effect, in which the transport of one photon prevents the transport of others, would be to separate two incoming indistinguishable photons to different output ports.
We show that time-energy uncertainty relations inherently prevent this ideal situation when the blockade is implemented by a two-level system. The fundamental nature of this limit is revealed in the fact that photon blockade in the strong coupling regime of cavity QED, resulting from the nonlinearity of the Jaynes-Cummings energy level structure, exhibits efficiency and temporal behavior identical to those of photon blockade in the bad cavity regime, where the underlying nonlinearity is that of the atom itself. We demonstrate that this limit can be exceeded, yet not avoided, by exploiting time-energy entanglement between the incident photons. Finally, we show how this limit can be circumvented completely by using a three-level atom coupled to a single-sided cavity, enabling an ideal and robust photon routing mechanism.

\end{abstract}
\maketitle
	
\section{\label{sec:level1}Introduction}

Often described as the optical equivalent of Coulomb blockade~\cite{fulton1987,kastner1992}, photon blockade~\cite{imamoglu1997}, in which the transport of only one photon through a nonlinear system is possible whereas excess photons are rejected, is one of the few photon-photon interactions that have been demonstrated experimentally to date~\cite{birnbaum2005,dayan2008,faraon2008}. Such nonlinear interactions at the single-photon level are inherently nonclassical, and form the basis for quantum information processing with photonic qubits~\cite{milburn1989,turchette1995a,schrader2004,duan2004}. In particular, the demonstration of the photon blockade is considered a significant step towards the goal of single photon routing~\cite{aoki2009}.
Two mechanisms of photon blockade have been demonstrated in recent years. Both mechanisms rely on the coupling of an atom (or another two-level system such as a quantum dot~\cite{young2010}) to cavity-enhanced electromagnetic modes.
In this work we establish that such an apparatus is inherently insufficient for the task of photon routing, and that a three-level system at least is needed to ensure a deterministic and efficient routing process. We demonstrate this by analyzing the simplest case of an input pulse containing exactly two photons, and deriving the probability for a successful routing event, namely that the system will respond to one photon differently than to the other.

The underlying conflict that limits the interaction of a two-level system with such a pulse, is the fact that the coupling to the electromagnetic mode dictates both the interaction bandwidth, and the memory time of the system. Thus, a pulse that is short enough to guarantee that the two photons arrive within the memory time of the system will have a bandwidth that exceeds the interaction bandwidth. Conversely, a pulse that is narrow-band enough to be included completely within the interaction bandwidth will be long enough to allow the system to respond to each photon as if it were the only one.

The outline of the paper is as follows. In Sec.~\ref{sec:level2} we present both analytical derivations and numerical calculations of the efficiency of two-photon routing based on photon blockade in cavity quantum electrodynamics (QED) \cite{miller2005,schuster2008}. The temporal behavior of the blockade mechanism is essentially the same in the bad cavity regime as in the strong coupling regime, despite the fact that the underlying nonlinearity is different in the two regimes. We show that the routing efficiency is inherently limited to~$\sim64\%$ in both cases. In Sec.~\ref{sec:level3} we study the possibility of exceeding this limit by using a time-energy entangled photon pair as the input pulse. This case seems especially relevant in light of the fact that such entanglement can be created by a two-level system coupled to a single electromagnetic mode~\cite{hofmann2003,kojima2003}, and so one could think of a two-stage process in which the first interaction creates the entanglement between the photons and the second interaction performs the routing mechanism. However, our analysis shows that even ideal time-energy entanglement can increase the routing efficiency only up to~$\sim77\%$. Time-energy entanglement generated by a two-stage interaction with a two-level system is even more limited and increases the efficiency only up to~$\sim68\%$.
Finally, in Sec.~\ref{sec:level4} we study the configuration of a three-level atom coupled to a single-sided cavity. Recent studies by Koshino \textit{et al.}~\cite{koshino2010} and Gea-Banacloche \textit{et al.}~\cite{gea-banacloche2011} have shown that this configuration enables deterministic mapping of a photonic state to the atom. We utilize this scheme to construct an ideal photon router that does not suffer from the inherent limits of photon blockade with a two-level system, and show that its efficiency can approach~$\sim100\%$ with realistic, experimentally achievable parameters.
\section{\label{sec:level2}Fundamental Limit for Photon Blockade with a Two-Level System}

The first demonstration of the photon blockade was performed with Cs atoms strongly coupled to a Fabry-P\'{e}rot cavity~\cite{birnbaum2005}, and later with quantum dots coupled to a photonic crystal resonator~\cite{faraon2008}. In this strong coupling regime, where the coupling rate between the atom and the cavity is larger than all the other rates in the system, the photon blockade relies on the anharmonicity of the energy levels of the coupled atom-cavity system~\cite{birnbaum2005,kubanek2008}, meaning that multi-photon excitations of the coupled system occur at different frequencies than a single-photon excitation. A similar anharmonicity of the spectrum of semiconductor systems has been used to demonstrate single-photon sources~\cite{michler2000,kim1999}.

The second photon blockade mechanism occurs at the bad cavity limit~\cite{rice1988}, in which the cavity-enhanced coupling of the atom to one electromagnetic mode is faster than its spontaneous emission to all other modes, yet slower than the cavity decay rate. Therefore, although strong coupling is not achieved, the atom interacts mostly with one electromagnetic mode and can thus be perceived as a one-dimensional atom~\cite{turchette1995b}. The blockade effect is then typically described as a dynamical process, equivalent to antibunching in free space resonance fluorescence~\cite{kimble1977}. This stems from the fact that in the bad cavity regime, the scattered photon escapes the cavity immediately, before the atom, which collapsed to the ground state, can rebuild its polarization and scatter another photon.
Nonetheless, one can consider this mechanism in the spectral domain as well, like in the strong coupling regime. The difference is that in this case the anharmonicity is of the $\it{atomic}$ energy levels, where, again, multi-photon excitations occur at different frequencies than a single-photon excitation. In either case, the final result can be described as a ``photon turnstile'' in continuous operation, since photons are transmitted (or reflected) one by one, as was demonstrated with Cs atoms coupled to a whispering gallery mode of a fiber-coupled microtoroid cavity~\cite{dayan2008,aoki2009}.

\subsection{Photon blockade in the bad cavity limit}
\begin{figure}[b]
\centering
\includegraphics[width=100mm,angle=0,scale=.9]{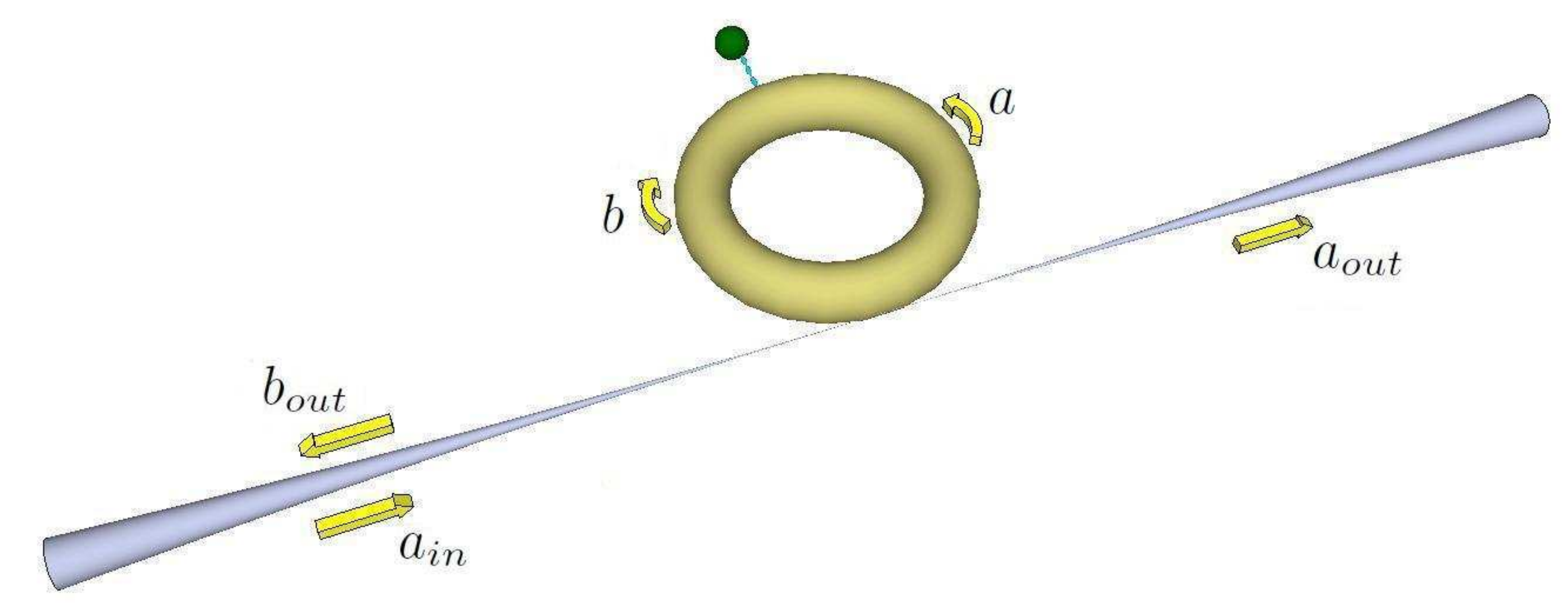}
\caption{\label{fig:setup}(Color online) Schematic depiction of a microtoroid setup. Note that $b_{in}$ (not shown here) is vacuum.}
\end{figure}

We begin by analyzing the blockade mechanism in the bad cavity limit. For this we consider a system similar to that of Ref.~\cite{aoki2009}, namely a single atom interacting with a fiber-coupled microtoroid, as depicted in Fig.~\ref{fig:setup}.

The two counterpropagating modes of the microtoroid, $\hat{a}$ and $\hat{b}$, are evanescently coupled to the atom with coupling strength $g$, and decay to the tapered fiber at a rate $2\kappa_{ex}\gg g$. Since we are interested in inherent limits to the routing effect, we shall assume at first that parasitic effects such as scattering between the cavity modes, intrinsic losses of the cavity and the natural decay of the atom are significantly slower than all the other processes in the system. The dominant parameter in the system is thus the cavity decay rate into the fiber, which brings us into the bad cavity regime. It is important, however, to make a distinction between this situation, in which the cavity is intentionally made to decay into the fiber, and a truly "bad" cavity that decays quickly due high intrinsic losses. Thus, we shall henceforth refer to this regime as the "fast cavity" regime, instead of the bad cavity regime.\\

We define the routing efficiency $C^{tr}$ as the probability of detecting one photon in the transmitted mode $\hat{a}_{out}$ and one in the reflected mode $\hat{b}_{out}$, given that the input mode $\hat{a}_{in}$ initially contains a two-photon pulse. Neither the time interval between the two clicks, nor their absolute time of occurrence are of importance, and hence, for calculating the probabilities for the four possible detection events, we integrate over all possible detection times:
\begin{eqnarray}
\label{Prr}
P^{tr}&=& \int_{0}^\infty\int_{-\infty}^\infty \Gamma^{tr}\,dtd\tau \\
P^{rt}&=& \int_{0}^\infty\int_{-\infty}^\infty \Gamma^{rt}\,dtd\tau \nonumber \\
P^{rr}&=& \int_{0}^\infty\int_{-\infty}^\infty \Gamma^{rr}\,dtd\tau \nonumber \\
P^{tt}&=& \int_{0}^\infty\int_{-\infty}^\infty \Gamma^{tt}\,dtd\tau \: , \nonumber
\end{eqnarray}
where $r$ stands for reflected and $t$ for transmitted, and
\begin{eqnarray}
\label{correlation1}
\Gamma^{tr}(t,\tau) &=& \langle \hat{a}^\dagger_{out}(t)  \hat{b}^\dagger_{out}(t+\tau)  \hat{b}_{out}(t+\tau)  \hat{a}_{out}(t)   \rangle \\
\Gamma^{rt}(t,\tau) &=& \langle \hat{b}^\dagger_{out}(t)  \hat{a}^\dagger_{out}(t+\tau)  \hat{a}_{out}(t+\tau)  \hat{b}_{out}(t)   \rangle  \nonumber\\
\Gamma^{rr}(t,\tau) &=& \langle \hat{b}^\dagger_{out}(t)  \hat{b}^\dagger_{out}(t+\tau)  \hat{b}_{out}(t+\tau)  \hat{b}_{out}(t)   \rangle  \nonumber\\
\Gamma^{tt}(t,\tau) &=& \langle \hat{a}^\dagger_{out}(t)  \hat{a}^\dagger_{out}(t+\tau)  \hat{a}_{out}(t+\tau)  \hat{a}_{out}(t)   \rangle \nonumber,
\end{eqnarray}
are the second order correlation functions of the two output modes. Note that this definition is already normalized so that $P^{tr}+P^{rt}+P^{rr}+P^{tt}=1$. The appropriate expression for the routing efficiency is thus given by
\begin{equation}
\label{definition}
C^{tr} = P^{rt}+P^{tr} .
\end{equation}
Accordingly, $P^{tt}$ and $P^{rr}$ denote the probabilities of the two failure mechanisms of the routing process, namely the probabilities for both photons to be transmitted, or for both photons to be reflected, respectively.

For an ideal router, events where both photons are transmitted or both reflected should not occur: $P^{tt}=P^{rr}=0$, and thus $C^{tr}=1$. In comparison, for a simple $50:50$ beam splitter, any of the four possibilities is equally likely, hence $C^{tr}=0.5$. Any useful photon router should thus satisfy $C^{tr}>0.5$, which is also the quantum limit, above which the Cauchy-Schwartz inequality for classical fields is violated.

\subsection{One- and two-photon source: the feeder cavity}
\begin{figure}[t]
\centering
\includegraphics[width=90mm,angle=0,scale=.9]{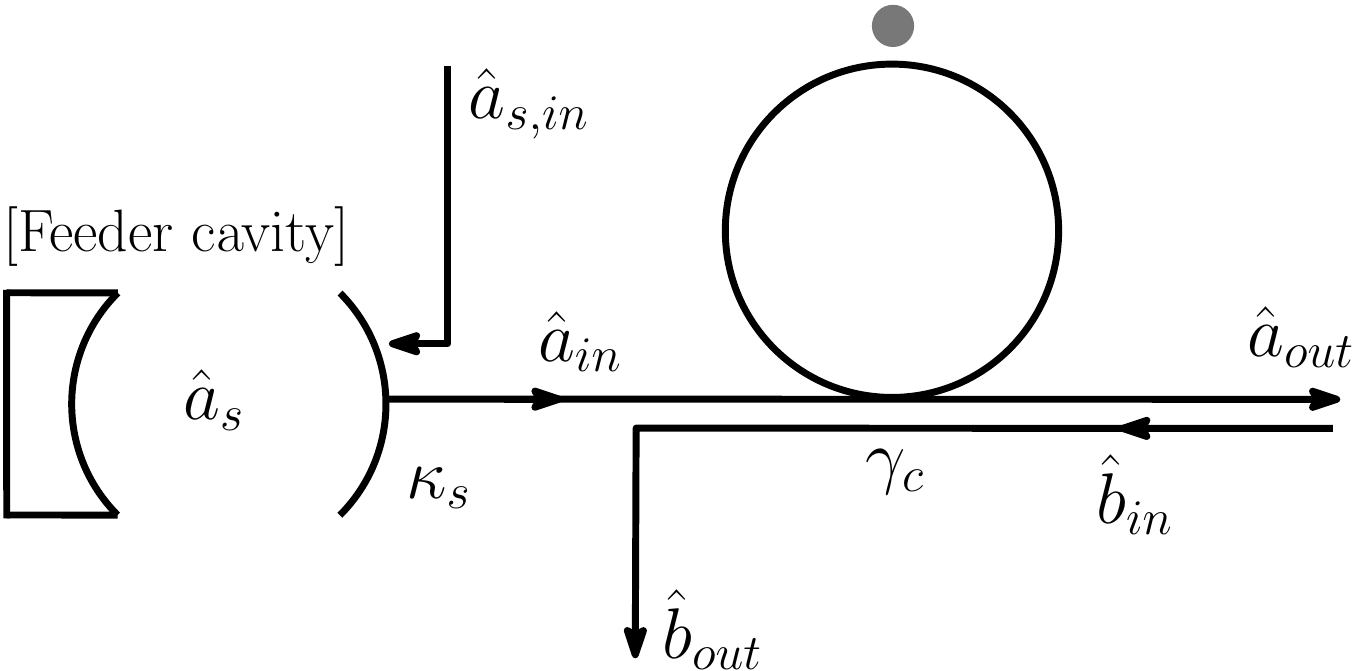}
\caption{\label{fig:feeder} A schematic of the photon router model. The input pulse is modeled by introducing a feeder cavity which leaks into the system with rate $2\kappa_{s}$.}
\end{figure}

In order to analyze the temporal behavior of a pulse interacting with this system and the resulting routing efficiency, we model the fiber by a feeder cavity containing the desired number of excitations (Fig.~\ref{fig:feeder}).

The decay rate of the feeder cavity $2\kappa_{s}$ gives a measure of the inverse of the resulting driving pulse width.
Using the input-output formalism~\cite{gardiner1985,gardiner1993,gardiner1994} and eliminating the microtoroid cavity modes adiabatically, which is justified in the fast cavity regime for which $\kappa_{ex}\gg g$, we can write for the output modes
\begin{subequations}\label{operators}
\begin{eqnarray}
\hat{a}_{out} &=& \sqrt{2\kappa_s}\hat{a}_s+ \sqrt{2\gamma_c}\hat{\sigma}\label{aout}\\
\hat{b}_{out} &=& \sqrt{2\gamma_c}\hat{\sigma}\label{bout},
\end{eqnarray}
\end{subequations}
where $\hat{a}_s$ is the annihilation operator of the feeder cavity field, $\hat{\sigma}$ is the lowering operator for the atom, and $2\gamma_c = 2g^2/\kappa_{ex}$ is the cavity-enhanced atomic decay rate per output channel. Since only normally-ordered moments of the output operators are considered, the vacuum noise input operators $\hat{a}_{s,in}$ and $\hat{b}_{in}$ have been discarded. Eq.~\eqref{aout} reflects the fact that emission of a photon into $\hat{a}_{out}$ by the feeder cavity and by the cavity-enhanced atom are indistinguishable. Accordingly, it is the interference between both paths that is detected. However, since we assumed the scattering between the cavity modes is negligible, only the atom is able to reflect a photon, and hence no interference will be observed in the reflected mode described in Eq.~\eqref{bout} (assuming $\hat{b}_{in}$ is vacuum).

The non-Hermitian effective Hamiltonian corresponding to the cascaded system of Fig.~\ref{fig:feeder} is given by ($\hbar=1$)~\cite{carmichael1993}
\begin{equation}
\label{hamiltonian1}
\hat{H_1}= -i\kappa_s \hat{a}_s^\dag \hat{a}_s -2i\gamma_c  \hat{\sigma}^\dag \hat{\sigma} - 2i\sqrt{\kappa_s\gamma_c}\hat{\sigma}^\dag\hat{a}_s.
\end{equation}
Note that the interaction between the feeder cavity and the system is unidirectional, and that it produces an exponentially decaying driving pulse. Also, since the atom can decay into two counterpropagating modes, the atomic population decay rate is $4\gamma_c$ instead of $2\gamma_c$.
\subsection{Results}
\subsubsection{Single photon driving pulse}
As a simple example, consider a driving pulse containing a single photon. The initial state is then
\begin{equation}
\left|\psi(0)\right\rangle=\left|1g\right\rangle,
\end{equation}
where the first index in the ket gives the number of photons in the feeder cavity, and the second index describes whether the atom is excited ($e$) or in its ground state ($g$). The state evolves according to the Schr\"{o}dinger equation $i\dot{\psi}=\hat{H_1}\psi$ to
\begin{equation}
\label{evolution}
\left|\psi(t)\right\rangle=a(t)\left|1g\right\rangle+b(t)\left|0e\right\rangle,
\end{equation}
with
\begin{eqnarray}
\label{solution1}
a(t) &=& e^{-\kappa_s t}\\
b(t) &=& -\frac{2\sqrt{\kappa_s\gamma_c}}{2\gamma_c-\kappa_s}\left(e^{-\kappa_s t}-e^{-2\gamma_c t}\right)\nonumber,
\end{eqnarray}
where henceforth the exponential functions are defined to be zero for $t<0$.
We use Eq.~\eqref{operators} to evaluate the photon flux at the output modes with this state, and integrate over time to define single-photon transmission and reflection probabilities
\begin{subequations}
\label{TR}
\begin{eqnarray}
T &=& \int_0^\infty{\langle \hat{a}_{out}^\dag(t)\hat{a}_{out}(t)\rangle\,dt} =  \frac{\kappa_s}{\kappa_s+2\gamma_c}  \label{T}\\
R &=& \int_0^\infty{\langle \hat{b}_{out}^\dag(t)\hat{b}_{out}(t)\rangle\,dt} =  \frac{2\gamma_c}{\kappa_s+2\gamma_c}  \label{R},
\end{eqnarray}
\end{subequations}
respectively.\\

As evident from Eq.~\eqref{TR}, for long enough pulses ($\kappa_s \ll \gamma_c$), $T\rightarrow0,R\rightarrow1$ and all light is reflected. This is due to the destructive interference between the atomic dipole radiation and the driving field in the forward direction. For short pulses ($\kappa_s \gg \gamma_c$), however, the atom has no time to build up the necessary dipole, and the photon is transmitted.

\subsubsection{Two photon driving pulse}
\begin{figure}[b!]
\centering
\includegraphics[width=90mm,angle=0,scale=0.9]{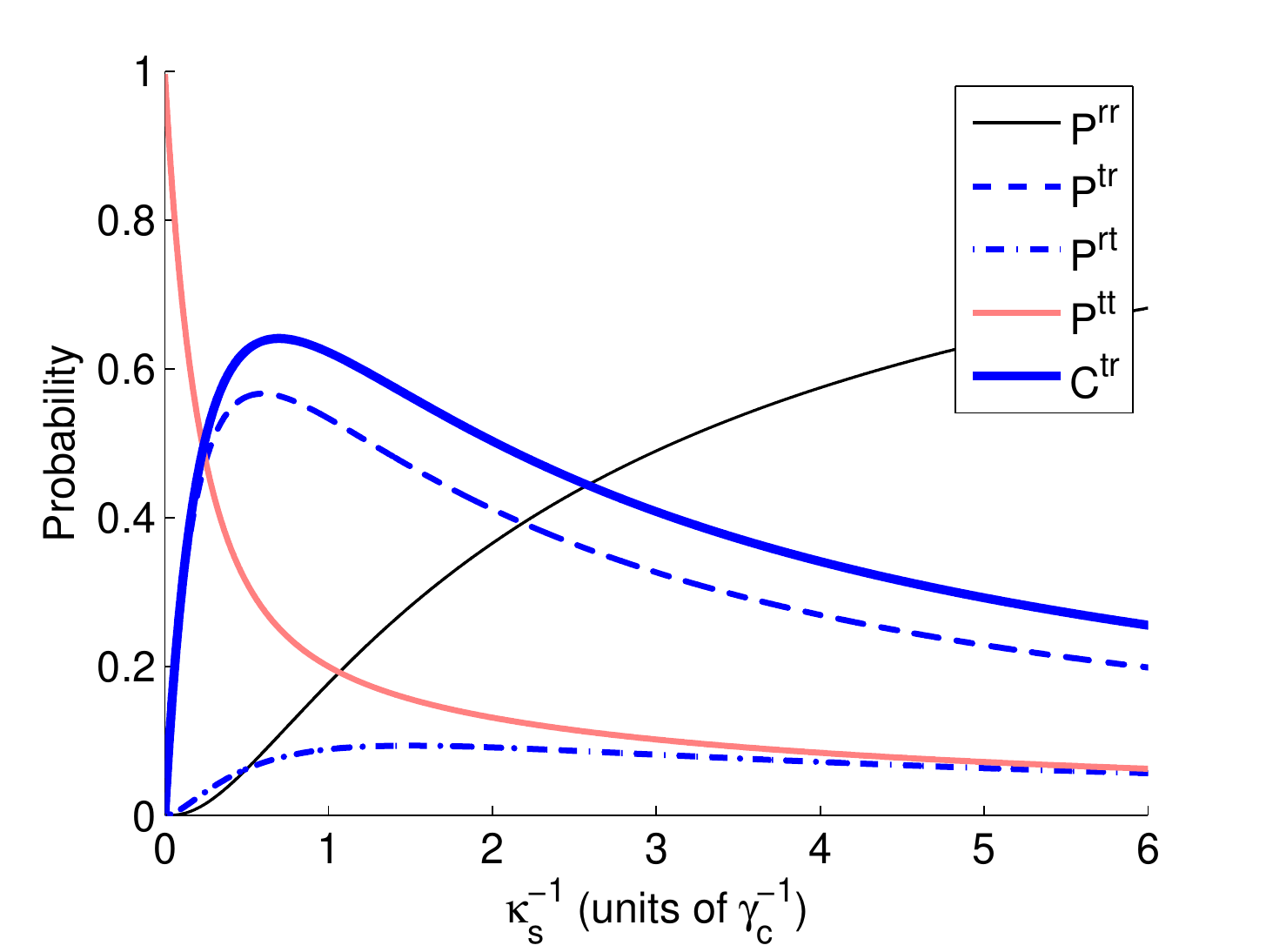}
\caption{\label{fig:turnstile}(Color online) Two-photon detection probabilities as a function of the pulse width $\kappa_s^{-1}$.}
\end{figure}

For deducing the routing properties of this system, we need to study the deviation from the linear behavior of Eq.~\eqref{TR} and analyze the case of more than one photon in the pulse. The source of the nonlinearity is evident already in Eq.~\eqref{operators}, in the presence of the atomic lowering operator in the expressions for the output mode operators. In particular, detection of a reflected photon projects the atom to its ground state, preventing the immediate scattering of a second photon and also disrupting the destructive interference in the forward direction, possibly allowing the second photon to slip through and be transmitted~\cite{carmichael1993}. In order to derive an analytic description for this effect we follow the evolution of a state in which the feeder cavity initially contains two photons. This initial state evolves to
\begin{equation}
\left|\psi(t)\right\rangle=\alpha(t)\left|2g\right\rangle+\beta(t)\left|1e\right\rangle,
\end{equation}
where
\begin{eqnarray}
\label{double-sided}
\alpha(t) &=& e^{-2\kappa_s t}   \\
\beta(t) &=&  -\frac{2\sqrt{2\gamma_c\kappa_s}}{2\gamma_c-\kappa_s}\left[e^{-\kappa_s t} - e^{-2\gamma_c t}\right]e^{-\kappa_s t}.\nonumber
\end{eqnarray}
After the detection of one photon, the wave function collapses to a state containing only a single excitation, which can be either in the cavity or in the atom. In the first case the subsequent evolution is again given by Eqs.~\eqref{evolution} and ~\eqref{solution1}, whereas in the second case, the state evolves to $\left|\psi(t)\right\rangle=c(t)\left|0e\right\rangle$, with
\begin{equation}
\label{solution2}
c(t) = e^{-2\gamma_c t}.
\end{equation}
Substituting these expressions into Eq.~\eqref{correlation1}, the second order correlation functions become
\begin{widetext}
\begin{eqnarray}
\label{correlation2}
\Gamma^{tr}(t,\tau) &=&4\left|\sqrt{\gamma_c}\left[ \left(\sqrt{2\kappa_s} \alpha(t) + \sqrt{\gamma_c}\beta(t)\right)b(\tau)+\sqrt{\kappa_s}\beta(t)c(\tau)\right]\right|^2 \\
\Gamma^{rt}(t,\tau) &=&4\left| \sqrt{\gamma_c}\beta(t)\left[\sqrt{\kappa_s}a(\tau)+\sqrt{\gamma_c}b(\tau)\right]\right|^2   \nonumber\\
\Gamma^{rr}(t,\tau) &=&4\left|\gamma_c\beta(t)b(\tau)\right|^2 \nonumber\\
\Gamma^{tt}(t,\tau) &=&4\left| \left[\sqrt{2\kappa_s} \alpha(t) + \sqrt{\gamma_c}\beta(t)\right]\left[\sqrt{\kappa_s}a(\tau)+\sqrt{\gamma_c}b(\tau)\right]
+\sqrt{\gamma_c \kappa_s}\beta(t)c(\tau)   \right|^2 \nonumber.
\end{eqnarray}
\end{widetext}
Substituting these correlation functions into Eq.~\eqref{Prr} and performing the integration leads to the following expressions for the various probabilities for transmission or reflection of the photons
\begin{eqnarray}
\label{Presults}
P^{tr}&=& \frac{12\kappa_s\gamma_c(\gamma_c+\kappa_s)}{(2\gamma_c+\kappa_s)^2(2\gamma_c+3\kappa_s)} \\
P^{rt}&=& \frac{4\kappa_s\gamma_c^2}{(2\gamma_c+\kappa_s)^2(2\gamma_c+3\kappa_s)}  \nonumber \\
P^{rr}&=& \frac{8\gamma_c^3}{(2\gamma_c+\kappa_s)^2(2\gamma_c+3\kappa_s)} \nonumber\\
P^{tt}&=& \frac{\kappa_s(3\kappa_s^2+4\gamma_c^2+2\kappa_s\gamma_c)}{(2\gamma_c+\kappa_s)^2(2\gamma_c+3\kappa_s)} \nonumber.
\end{eqnarray}
These probabilities and the corresponding routing efficiency $C^{tr}=P^{tr}+P^{rt}$ are presented in Fig.~\ref{fig:turnstile} as a function of the pulse width.

As evident, there is no pulse width for which both failure mechanisms $P^{tt},P^{rr}$ are zero, and so the routing efficiency $C^{tr}$ is a compromise, reaching a maximal value of only~$\sim64\%$. These results are consistent with previous works by Koshino \textit{et al.}~\cite{koshino2008,koshino2004} and Shapiro~\cite{shapiro2006}, although our model and derivation are different. The physical origin of the limited routing efficiency relies on the interplay between two counteracting effects. On one hand, if the pulse is significantly longer than $\gamma_c^{-1}$, the atom has time to reestablish its dipole after the first scattering, and therefore will be able to scatter the second photon backward as well when it arrives. On the other hand, due to time-energy uncertainty relations, a pulse that is significantly shorter than $\gamma_c^{-1}$ must have a bandwidth that exceeds $2\gamma_c$, which is also the interaction bandwidth of the atom. Thus, such a pulse will have some spectral components that cannot interact with the atom, and consequently there will be a probability for both photons to be transmitted. Since these two effects occur on the same timescale, the efficiency is inherently limited.
\begin{figure}[t!]
\centering
\includegraphics[width=90mm,angle=0,scale=1]{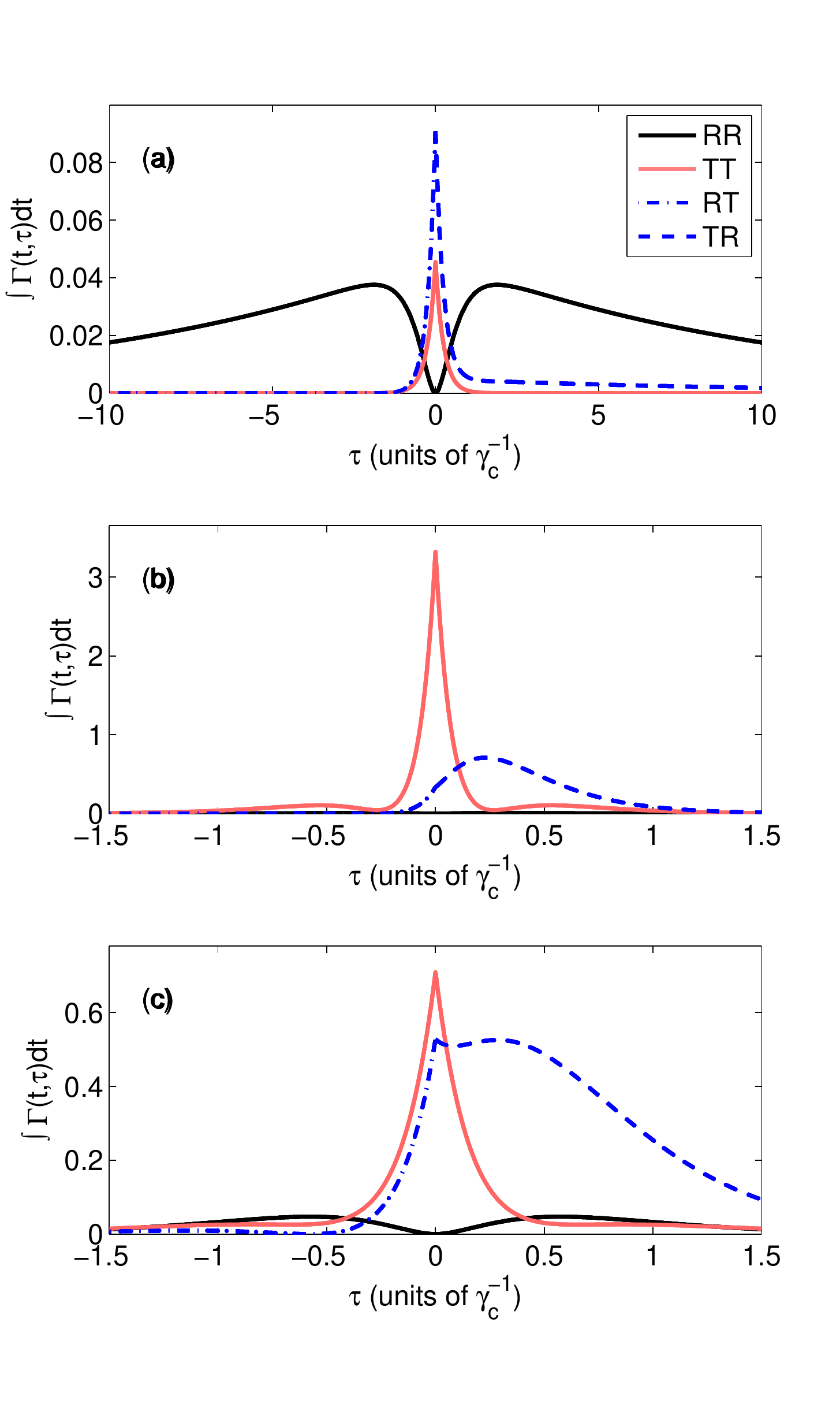}
\caption{\label{fig:gamma}(Color online) Marginal probability density functions of the various reflection and transmission events, as a function of the interval between the photon detection times $\tau$, for: (a) long pulses ($\kappa_s=0.05\gamma_c$), (b) short pulses ($\kappa_s=5\gamma_c$) and (c) intermediate pulses ($\kappa_s=1.5\gamma_c$). The different curves correspond to events of double reflection (RR), double transmission (TT), a transmission followed by a reflection (TR), and a reflection followed by a transmission (RT). Note that RR is practically zero in (b).}
\end{figure}

In order to gain more insight into the dynamics of this process, we present in Fig.~\ref{fig:gamma} the correlation functions of Eq.~\eqref{correlation2} with $t$ integrated over the entire pulse for intermediate, long and short pulses. This integration results in a marginal probability density function, displaying the relative probability for each of the four detection events as a function of the time interval between the two detections $\tau$. A second integration over $\tau$ would result in the probabilities of Eq.~\eqref{Presults}.\\

For all pulse lengths the reflection-reflection component RR is antibunched, corresponding to the projection of the atom to the ground state following the reflection of the first photon. This is the blockade effect, which implies that photons are reflected on a one-by-one basis. The antibunching of RR naturally leads to a complementary peak in the reflection-transmission graph RT, since at that timescale a second photon must be transmitted. The transmission-transmission component TT, however, results from different mechanisms at short and long pulses; at short pulses it reflects the probability that the (spectrally broad) photons did not interact with the atom at all, and so the peak in TT is as narrow as the pulse. For long pulses, TT results from the probability that the atom let one photon `slip through', since it was already excited at that moment by the other photon. In that case the atom can emit the second photon both forward and backward at equal probabilities; emission forward is the one that is presented in the TT graph, which is thus as wide as the atomic time scale $\gamma_c^{-1}$. Backward emission of the second photon leads to a transmission-reflection event TR, which is perhaps the most interesting process of the four. In particular, as we see in Fig.~\ref{fig:gamma}c, TR is the component that contributes the most to the routing efficiency $C^{tr}$ at intermediate pulses, and not, as one would have expected naively, the antibunching of the RR process, whose contribution is negligible at intermediate and short pulse lengths.\\
The underlying mechanism of the TR event with intermediate pulses is the simple fact that the photon emission rate from a feeder cavity occupied by two photons is twice as large as the emission rate from a cavity containing only one photon. Consequently, a first photon is very likely to arrive shortly after the beginning of the pulse, before the atom builds its dipole field, and thus it is likely to be transmitted. The time interval between the first photon and the arrival of the remaining second photon is expected to be longer, giving the atom enough time to reflect it. As in the case of TT, transmission of the first photon does not necessarily leave the second photon in the feeder cavity, but can also result in the collapse of the atom to its excited state, from which the photon has a 50\% probability of being reflected. These two mechanisms interfere constructively for the case of TR, but destructively for TT, thus considerably enhancing the routing efficiency.

\subsection{Simulations}
In our calculations so far, the temporal profile of the pulse is the only free parameter in the problem, and thus the limited efficiency of the photon blockade mechanism is inherent and unavoidable. In order to verify the analytic calculations, and also to check the effect of using various pulse shapes, a wave function approach using a nonunitary Hamiltonian was used to provide a fully quantized and complete simulation of the setup, including the microtoroid cavity (which was adiabatically eliminated in our analytical calculations) . The initial two-photon state is specified by
\begin{equation}
\left\vert \psi\right\rangle = \int\int_{-\infty}^\infty f(x_1,x_2)\frac{\hat{a}_r^\dagger(x_1)\hat{a}_r^\dagger(x_2)}{\sqrt2}\left\vert 0 g\right\rangle\,dx_1dx_2,
\end{equation}
where $\hat{a}_r^{\dagger}(x)$ and $\hat{a}_r(x)$ ($\hat{a}_l^{\dagger}(x)$ and $\hat{a}_l(x)$) create and annihilate a right (left) propagating photon at a location $x$ in the fiber. $\left\vert 0\right\rangle$ denotes the vacuum state of the microtoroid cavity, and $\left\vert g\right\rangle$ the ground state of the atom. $f(x_1,x_2)$ is a normalized weight function that describes the probability amplitude of the two photons to be located at $x_1$ and $x_2$. The effective Hamiltonian is given by~\cite{shen2009-I,shen2009-II}
\begin{align}
\label{Hamiltonian_sim}
H_{\text{eff}} = & \int dx \hat{a}_r^\dagger(x)\left(\omega_0-i v_g\frac{\partial}{\partial x}\right)\hat{a}_r(x)\\
&+                       \int dx \hat{a}_l^\dagger(x)\left(\omega_0+i v_g\frac{\partial}{\partial x}\right)\hat{a}_l(x)\nonumber\\
&+\left(\omega_c -i\kappa_i\right)\left( \hat{a}^\dagger \hat{a} + \hat{b}^\dagger \hat{b}\right)+\left(\omega_a-i\gamma\right)\hat{\sigma}_0^\dag\hat{\sigma}_0 \nonumber\\
&+  \int dx\, \chi(x)\left(V\hat{a}_r^{\dagger}(x) a + V^\ast\hat{a}^{\dagger}\hat{a}_r(x)\right)\nonumber\\
&+  \int dx\, \chi(x)\left(V\hat{a}_l^{\dagger}(x) b + V^\ast\hat{b}^{\dagger}\hat{a}_l(x)\right)\nonumber\\
&+ \left(g \hat{a}\hat{\sigma}_0^\dag + g^\ast \hat{a}^{\dagger}\hat{\sigma}_{0}\right) + \left(g^\ast \hat{b}\hat{\sigma}_0^\dag + g \hat{b}^{\dagger}\hat{\sigma}_{0}\right) \nonumber,
\end{align}

with $\hat{a}$ and $\hat{a}^{\dagger}$ ($\hat{b}$ and $\hat{b}^{\dagger}$) the annihilation and creation operators associated with the counterclockwise (clockwise) microtoroid cavity modes of frequency $\omega_c$. $\omega_0$ is the central frequency of the pulse around which the fiber dispersion relation is linearized. $\hat{\sigma}_0^\dag$ and $\hat{\sigma}_{0}$ are the bare atom raising and lowering operators of the transition with frequency $\omega_a$. $v_g$ is the group velocity of the pulse as it propagates through the fiber. $2\gamma$ is the population decay rate of the bare atom and $2\kappa_i$ is the intrinsic microtoroid cavity decay rate. The coupling strength between the fiber and the cavity is denoted by $V= \sqrt{2\kappa_{ex}v_g}$. The coupling between the cavity and the fiber is assumed to have a normal distribution $\chi(x)\propto e^{-\frac{1}{2}\left(2x/L_T\right)^2}$, where $L_T$ is the effective interaction length.

\begin{figure}[t!]
\centering
\includegraphics[width=90mm,angle=0,scale=0.9]{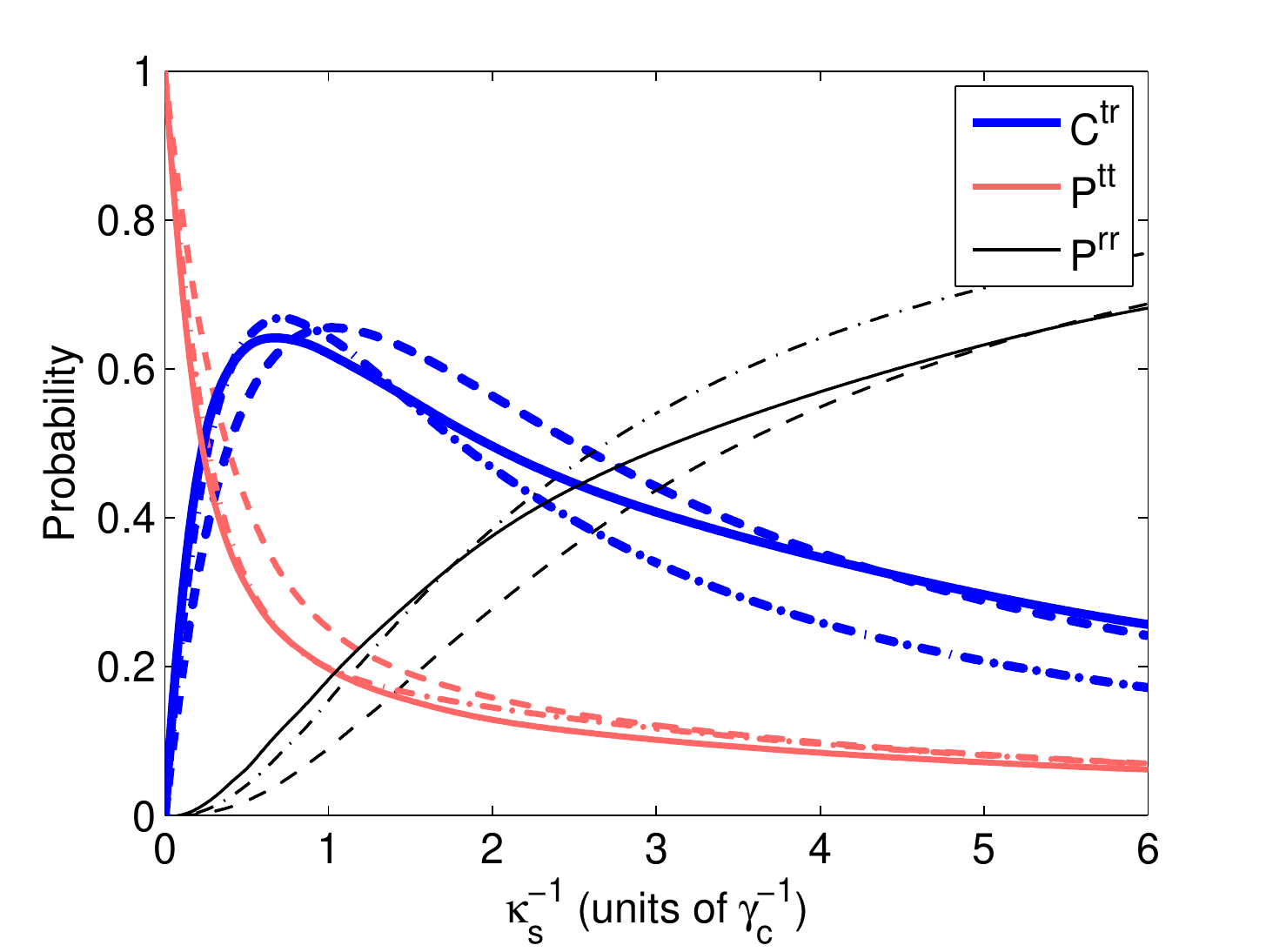}
\caption{\label{fig:shapes}(Color online) Simulated results for the two-photon detection probabilities as a function of the pulse width for square pulses (dashed curves), Gaussian pulses (dash-dotted curves) and exponential pulses (solid curves). For square and Gaussian pulses the pulse width $\kappa_s^{-1}$ is defined by their half width half maximum. The parameters are $g/2\pi=$ 70~MHz, $\kappa_{ex}/2\pi=$500~MHz, without losses ($\kappa_i=\gamma=0$). We note that the simulated results for exponential pulses are practically identical to the analytically calculated results presented in Fig. \ref{fig:turnstile}.}
\end{figure}

Figure~\ref{fig:shapes} presents the routing efficiency for various input pulse shapes, demonstrating little dependence on the pulse profile. Gaussian pulses have been found to yield the best results, with a routing efficiency of $C^{tr}=66.8\%$, yet the inherent conflict and the resulting limited efficiency remain.

\subsection{Photon blockade in the strong coupling regime}
The analysis so far relied on the fact that only one-photon transitions could take place, since atomic two-photon transitions (such as from $5S_{1/2}$ to $5D_{5/2}$ in Rb) occur at a different frequency. Thus, the atom behaves as a two-level system capable of dealing with only one excitation, with a linewidth dictated by the cavity-enhanced coupling to the output modes $\gamma_c$. Exactly the same results are expected in the case of photon blockade in the strong-coupling regime of cavity QED. In that regime, the nonlinear system is not the atom, but rather the coupled atom-cavity system, in which one-photon resonance and two-photon resonance are indeed at different frequencies~\cite{jaynes1963}. The linewidth of each of the vacuum Rabi sidebands in that case is $(\kappa+\gamma)/2$, with $\kappa$ being the bare cavity linewidth, and $\gamma$ the atomic linewidth in free-space. Thus, all our results so far are relevant for the photon blockade in the strong coupling regime as well, in which the interaction involves one of the Rabi sidebands, at detuning $g$ from the bare-cavity resonance. The only difference therefore is that the bandwidth $(\kappa+\gamma)/2$ replaces $\gamma_c$ in all the preceding expressions. We verified this by performing fully quantized simulations using the above method in the strong coupling regime, with the photons tuned to one of the Rabi sidebands. Indeed, these simulations yield results indistinguishable from those obtained in the fast cavity regime.\\

\section{\label{sec:level3} Routing Efficiency using Time-Energy Entangled Photons}

As established in the previous section, time-energy uncertainty relations are the key mechanism that limits the routing efficiency. Therefore, it is interesting to study the possible influence of using time-energy entangled photon pairs as the input two-photon pulse.
\subsection{Three-level atom source of photon pairs}
\begin{figure}[t!]
\centering
\includegraphics[width=90mm,angle=0,scale=0.9]{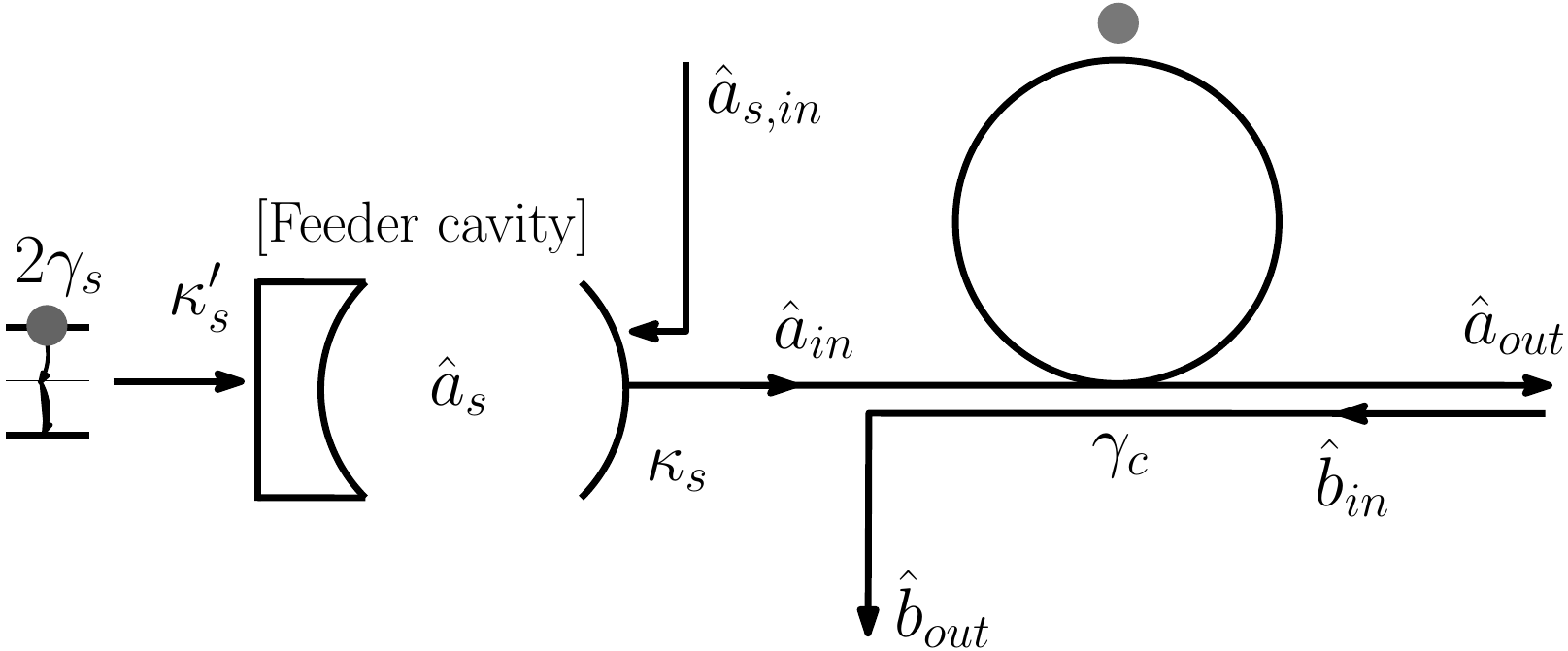}
\caption{\label{fig:feeder_entangled}The feeder cavity is driven by a three-level atom emitting two photons at once.}
\end{figure}

In order to obtain analytic expressions, the input pulse is now modeled by introducing a two-photon emitting atom with an infinitesimal intermediate state lifetime (Fig.~\ref{fig:feeder_entangled}) driving the feeder cavity. Two photons are thus simultaneously emitted into the feeder cavity at a rate $2\gamma_s$, from which they decay independently at a rate $2\kappa_s$. The feeder cavity then `smears out' the ideal entanglement generated by the three-level atom. Therefore, the ratio $\kappa_s/\gamma_s$ sets the amount by which the two-photon pulse that drives the router is temporally squeezed.

\begin{figure}[t!]
\centering
\includegraphics[width=90mm,angle=0,scale=0.9]{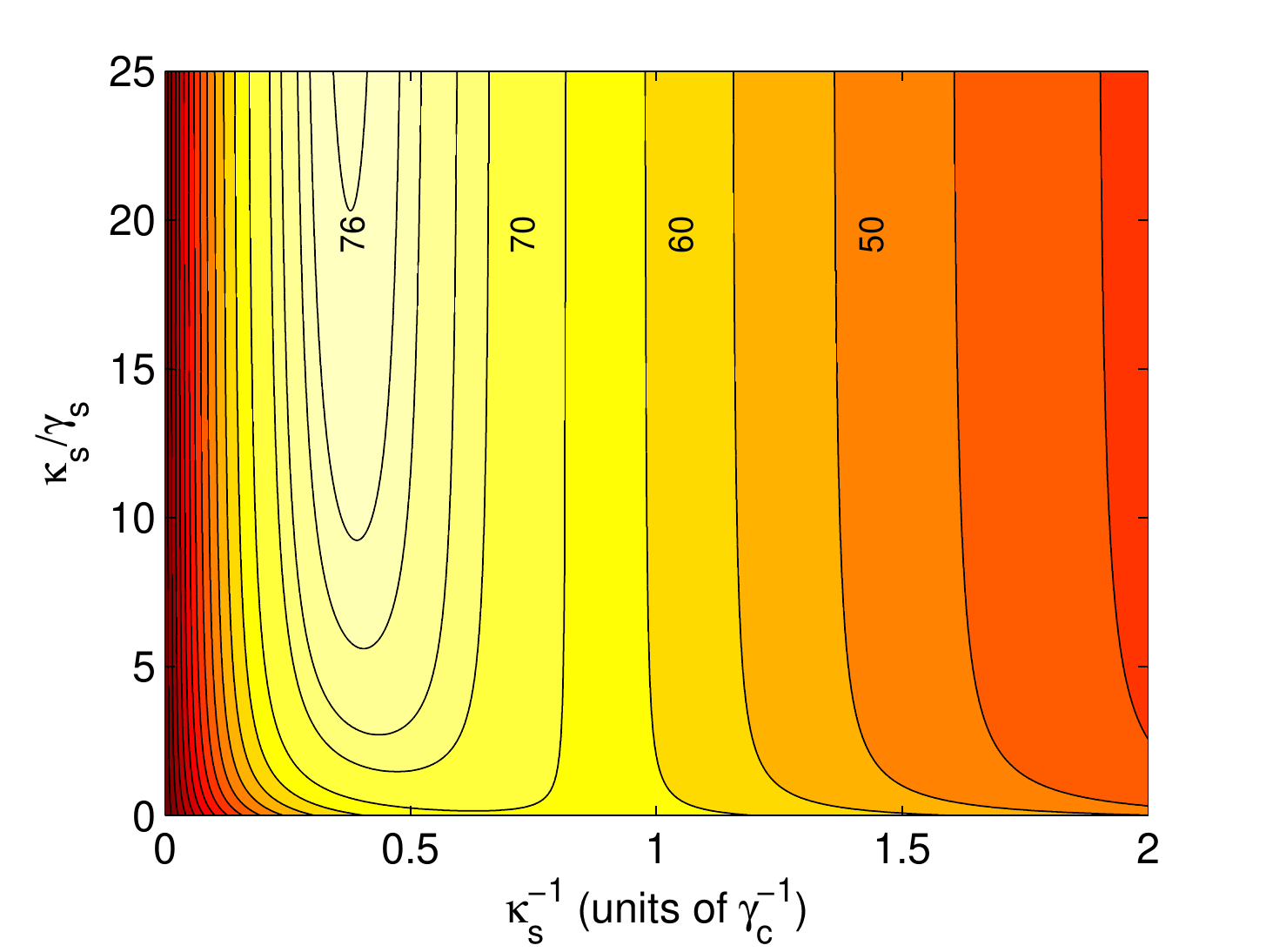}
\caption{\label{fig:gsgc}(Color online) The routing efficiency $C^{tr}$ as a function of the pulse length $\kappa_s^{-1}$ and the ratio $\kappa_s/\gamma_s$.}
\end{figure}

The effective Hamiltonian to be used in this case is
\begin{eqnarray}
\hat{H} &=& -i\kappa_s \hat{a}_s^\dag \hat{a}_s -2i\gamma_s  \hat{\sigma_s}^\dag \hat{\sigma_s}  -2i\gamma_c  \hat{\sigma}^\dag \hat{\sigma} \\
&&-2i\sqrt{2\kappa'_s\gamma_s}\hat{\sigma_s}\left(\hat{a}_s^\dag\right)^2 -  2i\sqrt{\kappa_s\gamma_c}\hat{\sigma}^\dag\hat{a}_s \nonumber,
\end{eqnarray}
where $\hat{\sigma_s}^\dag$, $\hat{\sigma_s}$ are the raising and lowering operators of the three-level atom for the transition from the upper state $e$ to the ground state $g$. $\kappa'_s\ll\kappa_s$ is the linewidth of the left mirror of the feeder cavity, and is taken to be very small but nonzero, to allow the feeder cavity to be driven by the three-level atom. Note that although the quantity of interest is the entangled pulse transmitted to the router, most of the pulse does not enter the feeder cavity, and hence only events whereby both photons enter the feeder cavity are taken into account. The two-photon state can be expanded as
\begin{equation}
\left|\psi(t)\right\rangle=\xi(t)\left|e0g\right\rangle + \alpha(t)\left|g2g\right\rangle+\beta(t)\left|g1e\right\rangle,
\end{equation}
where the first index describes the state of the three-level atom. Solving the Schr\"{o}dinger equation we obtain
\begin{widetext}
\begin{eqnarray}
\xi(t) &=& e^{-2\gamma_s t}   \\
\alpha(t) &=&  -2\frac{\sqrt{\kappa'_s\gamma_s}}{\kappa_s-\gamma_s}\left[e^{-2\gamma_s t} - e^{-2\kappa_s t}\right]\nonumber\\
\beta(t) &=& -4\sqrt{2\kappa'_s\kappa_s\gamma_s\gamma_c }\frac{(2\gamma_c-\kappa_s)e^{-2\gamma_s t} + (2\gamma_s-2\gamma_c-\kappa_s)e^{-2\kappa_s t}+ 2(\kappa_s-\gamma_s)e^{-(2\gamma_c+\kappa_s) t}}{(2\gamma_c-\kappa_s)(2\gamma_s-2\gamma_c-\kappa_s)(\kappa_s-\gamma_s)}.\nonumber
\end{eqnarray}
\end{widetext}
By substituting these expressions into Eq.~\eqref{correlation2}, the routing efficiency can be calculated using Eqs.~(\ref{Prr}-\ref{definition}), where $a(t)$, $b(t)$ and $c(t)$ are those of Eqs.~(\ref{solution1},\ref{solution2}). This leads to the following expressions for the transmission and reflection probabilities:
\begin{eqnarray}
\label{Presults2}
P^{tr}&=& \frac{4\kappa_s\gamma_c\left(6\gamma_s(\gamma_c+\kappa_s)+(\gamma_c+2\kappa_s)(2\gamma_c+3\kappa_s\right)}{(2\gamma_c+\kappa_s)^2(2\gamma_c+3\kappa_s)(2\gamma_s+2\gamma_c+\kappa_s)}\nonumber\\
P^{rt}&=& \frac{4\kappa_s\gamma_c^2\left(2(\gamma_s+\gamma_c)+3\kappa_s\right)}{(2\gamma_c+\kappa_s)^2(2\gamma_c+3\kappa_s)(2\gamma_s+2\gamma_c+\kappa_s)}\nonumber \\
P^{rr}&=& \frac{8\gamma_c^3\left(2(\gamma_s+\gamma_c)+3\kappa_s\right)}{(2\gamma_c+\kappa_s)^2(2\gamma_c+3\kappa_s)(2\gamma_s+2\gamma_c+\kappa_s)} \nonumber\\
P^{tt}&=& \frac{\kappa_s\left[\begin{array}{l}
8\gamma_c^2(\gamma_s+\gamma_c)+4\gamma_c\kappa_s(\gamma_s+2\gamma_c)\\
+2\kappa_s^2(3\gamma_s-2\gamma_c)+3\kappa_s^3
\end{array}\right]}
{(2\gamma_c+\kappa_s)^2(2\gamma_c+3\kappa_s)(2\gamma_s+2\gamma_c+\kappa_s)}.
\end{eqnarray}
Note that these equations coincide with Eqs.~\eqref{Presults} in the limit of an unentangled input pulse ($\gamma_s\rightarrow\infty$). We see that the routing efficiency (Fig.~\ref{fig:gsgc}) increases significantly as the entanglement is increased, reaching a 10\% improvement at $\kappa_s/\gamma_s\approx5$. For even larger values of $\kappa_s/\gamma_s$, the increase becomes less pronounced, and the efficiency tends to an asymptotic value of~$\sim77\%$. Thus, we see that even an entangled state at the input cannot completely eliminate the conflict that limits the routing efficiency of a two-level system.

\begin{figure}[b!]
\centering
\includegraphics[width=90mm,angle=0,scale=0.9]{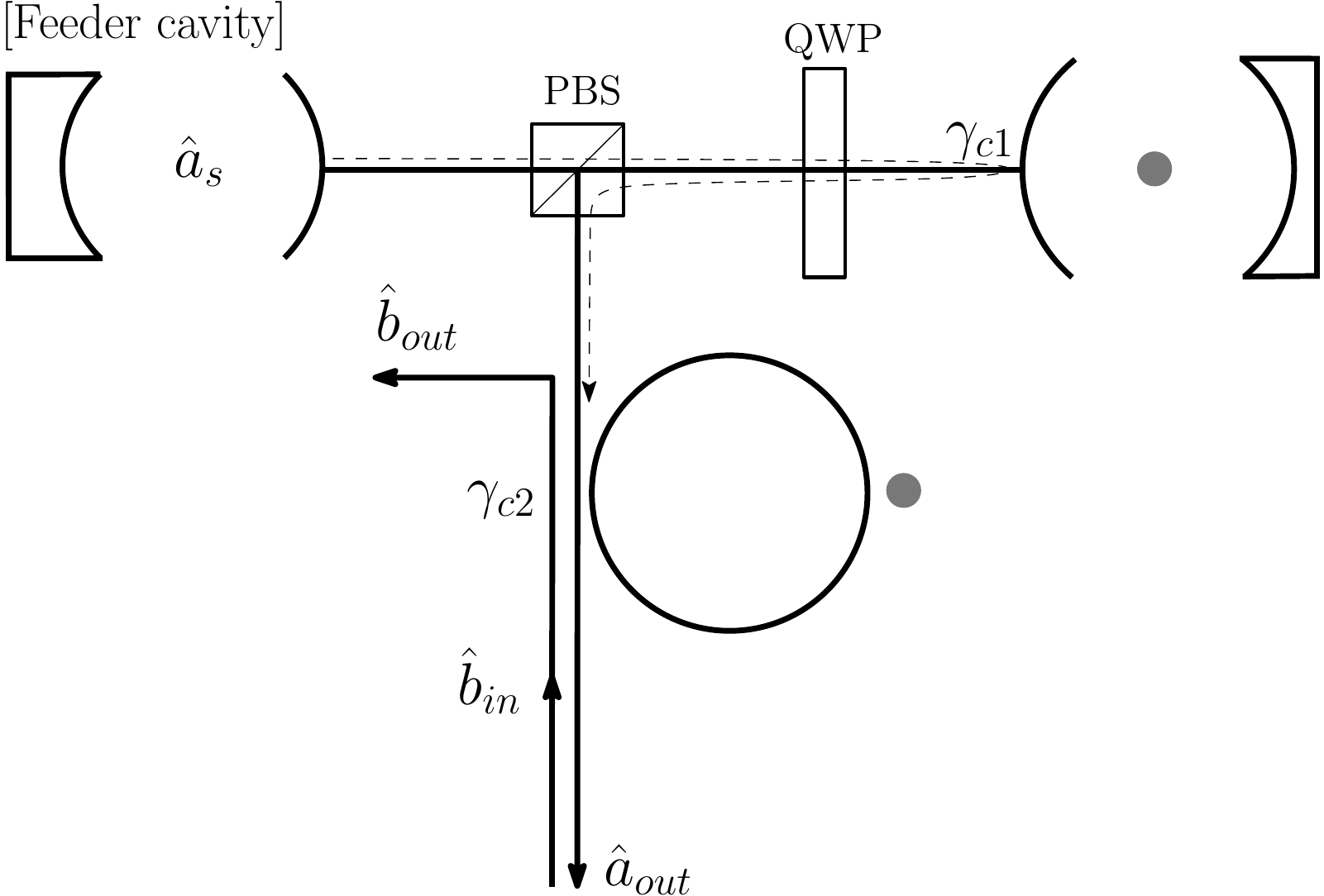}
\caption{\label{fig:cascaded}Cascading of a single-sided cavity QED system and a photon router. Due to the quarter wave plate (QWP), the pulse is reflected by the polarizing beam splitter (PBS) after interaction with the single-sided cavity. The trajectory of the two-photon pulse is shown by the dashed line.}
\end{figure}
\subsection{Cavity QED source of photon pairs}

Although an entangled two-photon input pulse does not enable ideal routing, the increased efficiency obtained in this scenario seems potentially useful, especially since time-energy entanglement can be generated within a two-photon pulse by reflecting it from a single-sided cavity containing a two-level atom~\cite{hofmann2003,kojima2003}, as depicted in Fig.~\ref{fig:cascaded}. Thus, one could imagine performing a two-stage interaction with a cavity QED system, the first for generating the entanglement, and the second for performing the routing.
The underlying physical effect behind the generation of entanglement is the same interference that occurs between the driving field and the atomic dipole, which suppresses the forward transmission in the regular cavity configuration, and leads to the backscattering of the photon. However, in the case of a single-sided configuration, there is only one cavity mode, and the destructive interference between the atomic dipole and the driving field is only partial, such that the emitted photon flux equals the incoming photon flux. Once a photon is scattered, however, the atomic state flips its phase, and subsequently the destructive interference becomes a constructive one. In more detail, the effective Hamiltonian modeling the passage of a pulse through a single-sided cavity QED system in the fast cavity regime is given by
\begin{equation}
\label{hamiltonian2}
\hat{H_2}= -i\kappa_s \hat{a}_s^\dag \hat{a}_s -i\gamma_{c1}  \hat{\sigma}^\dag \hat{\sigma} - 2i\sqrt{\kappa_s\gamma_{c1}}\hat{\sigma}^\dag\hat{a}_s,
\end{equation}
where we assume that the atom is coupled to only one standing wave in the cavity, explaining why the decay rate of the atom is twice as small as in Eq.~\eqref{hamiltonian1}.
The two-photon state is now expanded as
\begin{equation}\label{bunched_state}
\left|\psi(t)\right\rangle=\alpha(t)\left|2g\right\rangle+\beta(t)\left|1e\right\rangle,
\end{equation}
with
\begin{eqnarray}
\alpha(t) &=& e^{-2\kappa_s t}\nonumber\\
\beta(t) &=&  -2\sqrt{2}\frac{\sqrt{\gamma_{c1}\kappa_s}}{\gamma_{c1}-\kappa_s}\left[e^{-\kappa_s t} - e^{-\gamma_{c1} t}\right]e^{-\kappa_s t}.
\end{eqnarray}
Note that due to the single-sided configuration of the cavity, these expressions differ from those of Eq.~\eqref{double-sided}. It is evident that the driving and the atomic radiation are out of phase. To illustrate the dynamics more clearly let us assume a long pulse ($\kappa_s\ll\gamma_{c1}$) and neglect transient effects that occur at $t\sim\gamma_{c1}^{-1}$. Under these approximations the state of the system is
\begin{equation}
\label{state1}
\left|\psi(t)\right\rangle \approx e^{-2\kappa_s t} \left[ \left|2g\right\rangle-2\sqrt{2}\sqrt{\frac{\kappa_s}{\gamma_{c1}}}\left|1e\right\rangle \right] \:.
\end{equation}
Applying the output operator of Eq.~\eqref{aout} yields
\begin{equation}
\hat{a}_{out}\left|\psi(t)\right\rangle\approx -2\sqrt{\kappa_s} e^{-2\kappa_s t}\left[ \left|1g\right\rangle+2\sqrt{\frac{\kappa_s}{\gamma_{c1}}}\left|0e\right\rangle \right] \:,
\end{equation}
corresponding to a photon detection probability rate of $ \left\langle \: \hat{a}_{out}^\dag \hat{a}_{out} \right\rangle \approx 4 \kappa_s \: e^{-4\kappa_s t}$. As evident, the detection of the first photon has led to a sign flip between the driving field and the atom, resulting now in a constructive interference, rather than the previous destructive one. Normalizing the state and applying $\hat{a}_{out}$ again shows that this leads to a sudden increase of the photon detection rate by a factor of $4.5$ to $ \left\langle \: \hat{a}_{out}^\dag \hat{a}_{out} \right\rangle \approx 18 \: \kappa_s e^{-4\kappa_st}$. Since after one photon detection there is only one excitation left in the system, naively one could have expected the probability for a second photon detection to drop and not increase; this increase is therefore in fact by a factor of 9 compared to the expected value at steady state, and thus the output state exhibits strong bunching and consequently time-energy entanglement between the two photons~\cite{kojima2003}.\\

\begin{figure}[t!]
\centering
\includegraphics[width=90mm,angle=0,scale=0.9]{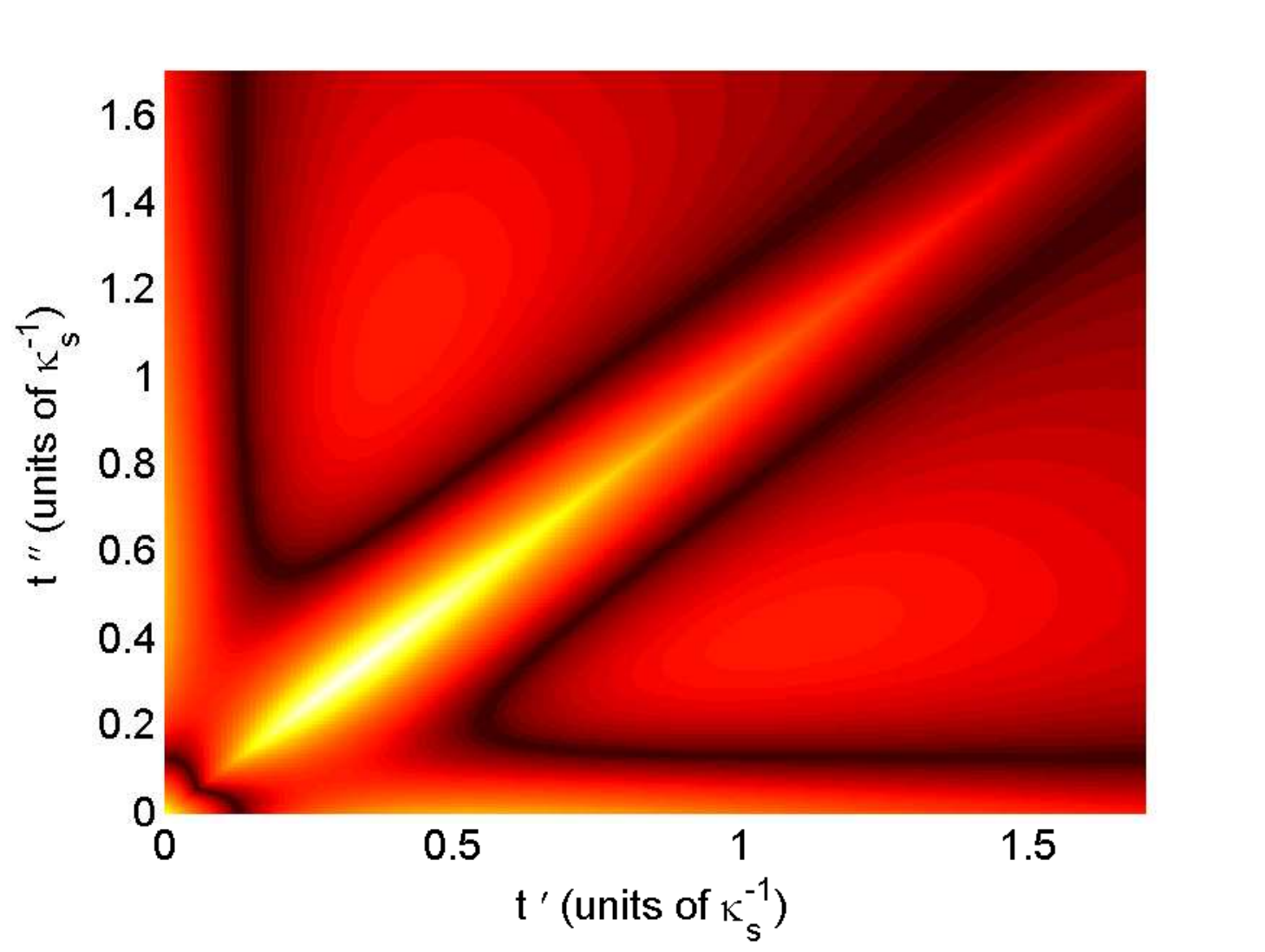}
\caption{\label{fig:single-sided}(Color online) Probability of detecting two photons at times $t'$ and $t''$ for an exponential pulse ($\gamma_{c1} = 5\kappa_s$) after passage through a single-sided cavity QED system.}
\end{figure}

Solving the evolution of the two-photon state \eqref{bunched_state} using Eq.~\eqref{hamiltonian2} without approximations, we get that the probability amplitude of photon detections at times $t$ and $t+\tau$ is (Fig.~\ref{fig:single-sided})
\begin{eqnarray}
f(t,\tau) &=& \left[1-4e^{-(\gamma_{c1}-\kappa_s)\tau}\right]e^{-2\kappa_s t}e^{-\kappa_s \tau}\\
&&-2\left[1-3e^{-(\gamma_{c1}-\kappa_s)\tau}\right]e^{-\gamma_{c1} t} e^{ -\kappa_s t}e^{-\kappa_s \tau}. \nonumber
\end{eqnarray}
The terms in the second brackets correspond to the transient behavior of the cavity-enhanced atom, and have influence only for short pulses, as evident from their fast decay by $e^{-\gamma_{c1} t}$. The two terms in the first brackets thus demonstrate the main dynamics of the system. The first term presents the possibility that no interaction occurred with the atom, and the second, nonlinear term is the one that leads to the bunching of the two-photon pulse. \\

The entanglement appearing in Fig.~\ref{fig:single-sided} indeed suggests that by utilizing this pulse as the input to a routing system, higher efficiency might be obtained. In order to investigate this possibility, a triply cascaded system needs to be considered (Fig.~\ref{fig:cascaded}). The feeder cavity drives a single-sided cavity QED system with effective linewidth $\gamma_{c1}$, and the resulting bunched light drives the photon router with linewidth $\gamma_{c2}$.\\

\begin{figure}[t!]
\centering
\includegraphics[width=90mm,angle=0,scale=0.9]{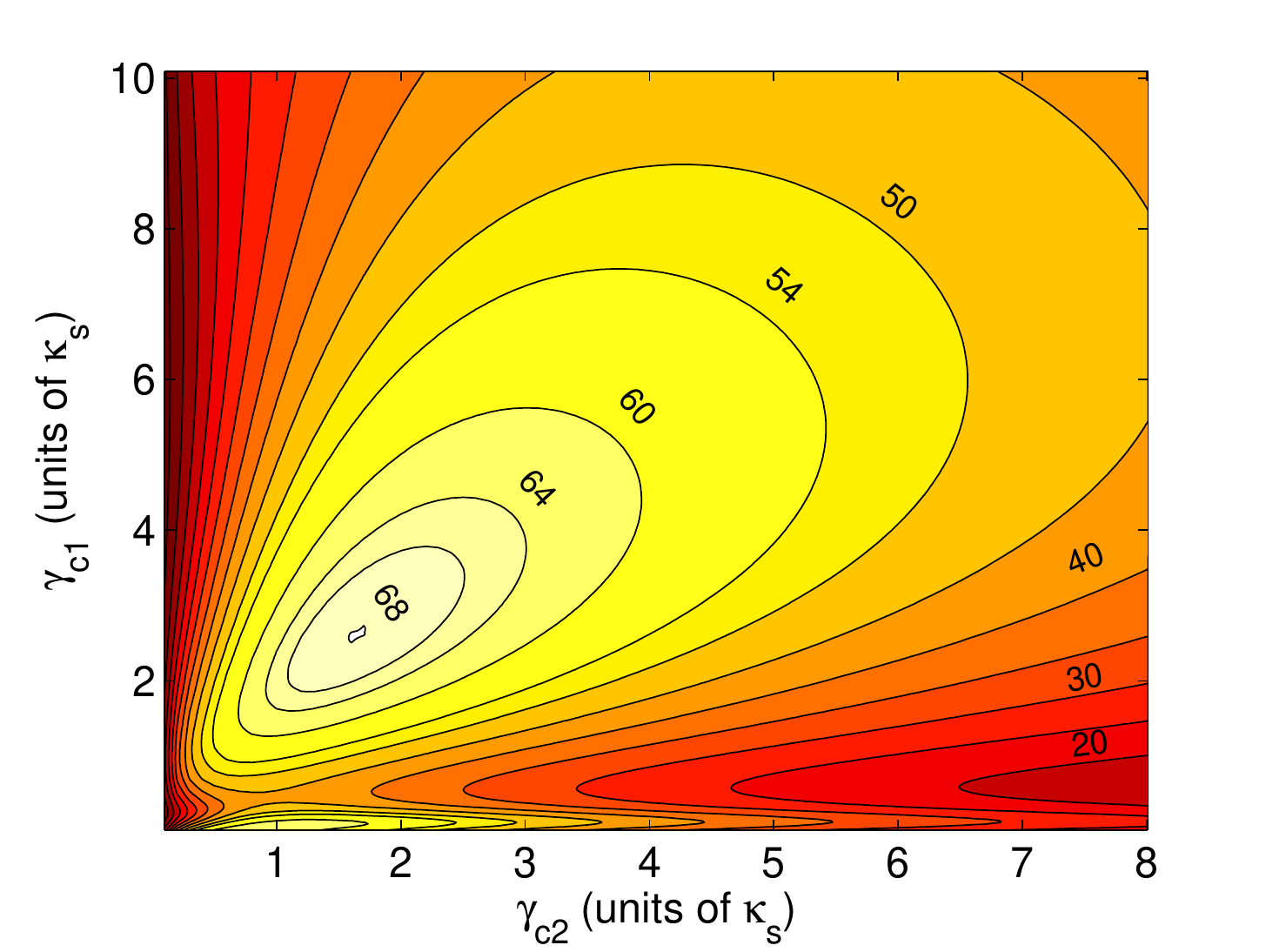}
\caption{\label{fig:gc1gc2}(Color online) Routing efficiency as a function of the cavity-enhanced decay rates during the first and second passages through the system.}
\end{figure}

The Hamilonian of this system consists of the Hamiltonians of its subsystems, given by Eq.~\eqref{hamiltonian1} and Eq.~\eqref{hamiltonian2}, and a term corresponding to the driving of the photon router by the single-sided cavity
\begin{equation}
\label{hamiltonian3}
\hat{H_{3}}= \hat{H_{1}}+\hat{H_{2}}-2i\sqrt{\gamma_{c2}}\hat{\sigma}_2^\dag(\sqrt{\kappa_s}\hat{a}_s+\sqrt{\gamma_{c1}}\hat{\sigma}_1).
\end{equation}
The analytic solution of the detection probabilities is calculated in the appendix. The resulting routing efficiency is shown in Fig.~\ref{fig:gc1gc2}. \\

Note that for $\gamma_{c1}=0$ we recover the result of Fig.~\ref{fig:turnstile}. By increasing $\gamma_{c1}$ the pulse becomes bunched, and the routing efficiency reaches~$\sim 68\%$. However, while the bunching ratio increases, the bunching efficiency decreases. This is due to the decreasing area of the bunched part in Fig.~\ref{fig:single-sided} as it becomes narrower. Hence, further increase of $\gamma_{c1}$ ceases to improve the routing efficiency. We conclude that a two-stage interaction with the cavity-enhanced atom, can only improve the routing efficiency by at most~$\sim 4\%$.

\section{\label{sec:level4}Ideal routing Efficiency using a Three-level Atom}

\begin{figure}[b]
\centering
\includegraphics[width=90mm,angle=0,scale=0.9]{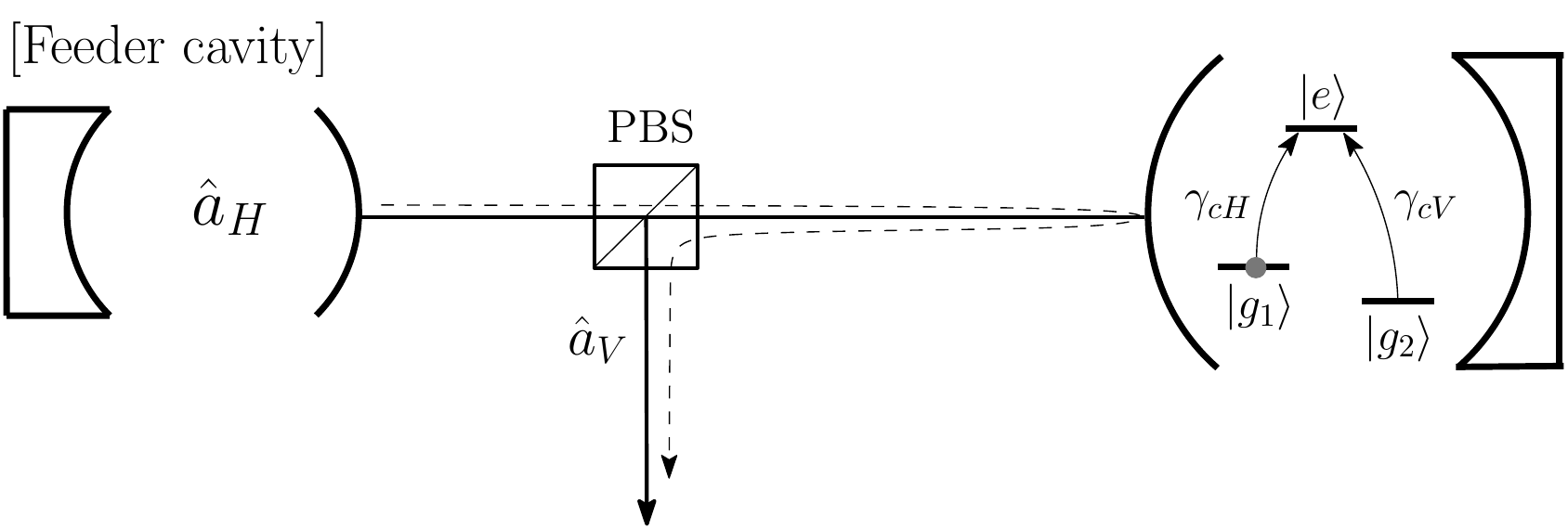}
\caption{\label{fig:three}Schematic depiction of a three-level atom in the $\Lambda$-configuration inside a single-sided cavity. The trajectory of the pulse is shown by the dashed line.}
\end{figure}

In the previous sections we have studied the limitations of a two-level system as a photon router. The situation changes dramatically when another cavity-enhanced atomic transition is introduced, for example by utilizing a three-level atom in the $\Lambda$-configuration inside the cavity. We shall consider the behavior of this system in the single-sided cavity setup, as depicted in Fig.~\ref{fig:three}.

The two ground states are denoted by $|g_1\rangle$ and $|g_2\rangle$, and the excited state by $|e\rangle$. We denote the fields resonant with the $g_1\rightarrow e$ transition and with the $g_2\rightarrow e$ transition as $H$ and $V$ fields, respectively, assuming they differ in polarization. This enables separation of both fields by a polarizing beam-splitter (PBS). In this configuration, therefore, routing to the two different output ports is performed by manipulating the polarization of the photons
\endnote{Alternatively, the modes can differ in their frequency. In that case, spatial separation is obtained by using a frequency filter such as a Bragg grating.}.\\
Taking the feeder cavity to be resonant with the $H$-transition, the appropriate Hamiltonian is

\begin{equation}
\label{hamiltonian4}
\hat{H}= -i\kappa_s \hat{a}_H^\dag \hat{a}_H -i\left(\gamma_{cH}+\gamma_{cV}\right)\hat{\sigma}_{ee} - 2i\sqrt{\kappa_s\gamma_{cH}}\hat{\sigma}_{1e}\hat{a}_H,
\end{equation}

where $\hat{\sigma}_{ee}=|e\rangle \langle e|$ and $\hat{\sigma}_{1e}=|g_1\rangle \langle e|$. $2\gamma_{cH}$, $2\gamma_{cV}$ are the cavity-enhanced decay rates from the excited state into ground states $|g_1\rangle$ and $|g_2\rangle$, respectively.

For a long input pulse ($\kappa_s\ll\gamma_{cH},\gamma_{cV}$) containing a single $H$-photon, the probability of a full transfer to the right ground state $|g_2\rangle$ starting with an atom in the left ground state $|g_1\rangle$ is given by
\begin{equation}
P_V= \frac{1}{1+\left(\gamma_{cV}-\gamma_{cH} \right)^2/4\gamma_{cV}\gamma_{cH}}.
\end{equation}
Thus, by choosing transitions for which $\gamma_{cH}\simeq\gamma_{cV}=\gamma_{c}$, one can approach $P_V\rightarrow 1$, and obtain deterministic transfer of the atom from one ground state to the other. In that process, the $H$-photon is absorbed and released as a $V$-photon. This configuration was studied by Koshino \textit{et al.}, who demonstrated that it may be used to implement deterministic quantum state transfer and a $\sqrt{\textrm{SWAP}}$ gate between a photon and an atom~\cite{koshino2010,hong2008}.

In order to use this apparatus for photon routing, we generalize this result by starting with a Fock state $|n\rangle$ of type $H$ in the feeder cavity. Assuming the atom is initialized in the left ground state $|g_1\rangle$, the wave function evolves to
\begin{equation}
\left|\psi(t)\right\rangle=\alpha(t)\left|n,g_1\right\rangle+\beta(t)\left|n-1,e\right\rangle. \:
\end{equation}
By solving the Schr\"{o}dinger equation with the Hamiltonian of Eq.~\eqref{hamiltonian4}, we get
\begin{eqnarray}
\label{three}
\alpha(t)&=& e^{-n\kappa_s t}\\
\beta(t) &=& -\frac{2\sqrt{n\kappa_s\gamma_{cH}}}{\gamma_{cH}+\gamma_{cV}-\kappa_s} \left(   e^{-n\kappa_s t} -   e^{-[(n-1)\kappa_s +\gamma_{cH}+\gamma_{cV} ]t}    \right)\nonumber.
\end{eqnarray}
Using the output operators from Eq.~\eqref{operators}, we get that the probability of first detecting a $V$-photon is
\begin{widetext}
\begin{equation}
P_V(n)= \frac{1}{1+\left[\left(\gamma_{cV}-\gamma_{cH}\right)^2+\left(\gamma_{cV}+\gamma_{cH}\right)\left(3n-2 \right)\kappa_s + \left(2n-1 \right)\left(n-1\right)\kappa_s^2\right]/4\gamma_{cH}\gamma_{cV}}.
\end{equation}
\end{widetext}

Once again, by using a long enough pulse and symmetric transitions, the first detected photon will with certainty be a $V$-photon. When that happens, the atom collapses to the ground state $|g_2\rangle$, and becomes transparent to all the remaining $H$-photons, which are consequently reflected from the cavity unchanged. The mapping resulting from the passage through the system is thus
\begin{equation}
\left(\hat{a}_H^\dag\right)^n|0\rangle\rightarrow\left(\hat{a}_H^\dag\right)^{n-1}\hat{a}_V^\dag|0\rangle,
\end{equation}
\begin{figure}[b]
\centering
\includegraphics[width=90mm,angle=0,scale=0.9]{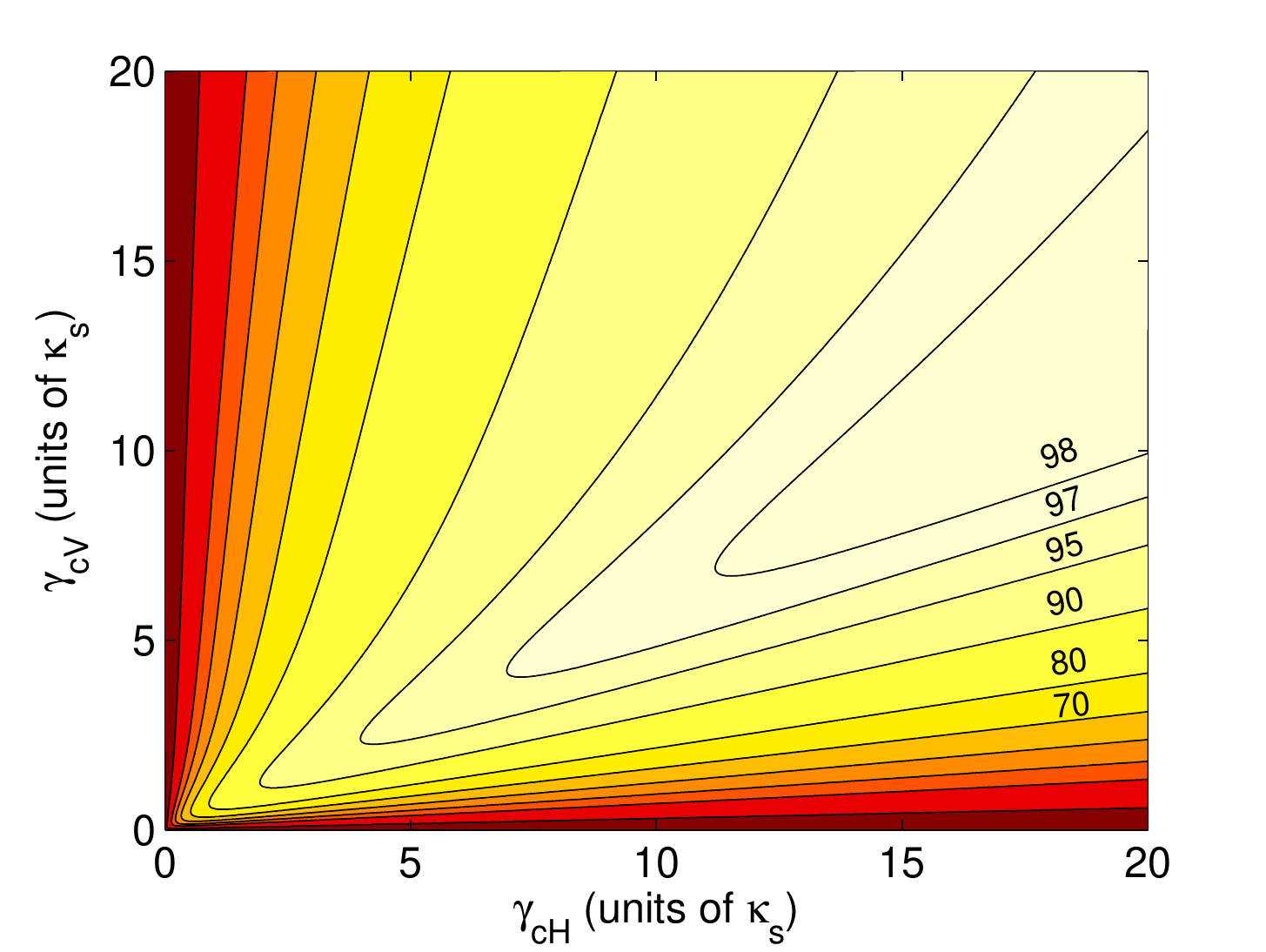}
\caption{\label{fig:gchgcv}(Color online) Routing efficiency $C^{tr}$ as a function of the cavity-enhanced decay rates $\gamma_{cH}$ and $\gamma_{cV}$.}
\end{figure}
i.e., it acts as an ideal photon router, which is transparent to all but one photon.

Specifically, in the case of $n=2$, we obtain for the routing efficiency (Fig.~\ref{fig:gchgcv})
\begin{equation}
C^{tr}=P_{VH}+P_{HV},
\end{equation}
where $P_{VH}=P_V(n=2)$ is the probability of detecting an $H$-photon after a $V$-photon has been detected, and $P_{HV}$ is the probability of first detecting an $H$-photon and then a $V$-photon. Taking into account that $P_{VH}=P_V(n=2)$ can be made close to unity by itself, we see that this routing process includes an inherent robustness, or error-correction mechanism, since in the unlikely event in which the first photon stays $H$ (for example, due to a slight mismatch between $\gamma_{cH}$ and $\gamma_{cV}$ or due to non-negligible $\kappa_s$), the second photon is likely to be turned into $V$, leading to the desired result nonetheless. Naively, one could expect $P_{HV}$ to be $(1-P_{VH})P_V(n=1)$, namely the probability that the first photon was not turned to $V$ times the probability of turning the last remaining photon into $V$, however $P_V(n)$ is correct only for an atom that begins at the ground state $|g_1\rangle$, whereas after detection of an $H$ photon the atom has some probability amplitude to be in the excited state $|e\rangle$. The full solution for $P_{HV}$, following the same procedure as for $P_{VH}$, is:
\begin{equation}
P_{HV}= \frac{\kappa_s(7\gamma_{cV}-\gamma_{cH}+6\kappa_s)+ \left(\gamma_{cV}-\gamma_{cH}\right)^2}{(\kappa_s+\gamma_{cH}+\gamma_{cV})(\gamma_{cH}+\gamma_{cV})}P_V(n=2).
\end{equation}
The overall routing efficiency is presented in Fig.~\ref{fig:gchgcv}. The contribution of $P_{HV}$ adds to already-high values of $P_{VH}$ to result in efficiencies that approach unity for reasonable $\gamma_{cV}$,$\gamma_{cH}$, and thus we see that this process acts as a robust and nearly ideal photon router, free from the limitations discussed in the previous sections.\\

Finally, we wish to estimate the robustness of this routing scheme to losses in the system. We approximate the effect of the coupling to the environment by calculating the probability of a quantum jump in the system, neglecting the change in Eq.~\eqref{three}. The effect of spontaneous emission into the vacuum at rate $\gamma$ for long pulses and $\gamma_{cH} =\gamma_{cV}=\gamma_{c}$ is then given by
\begin{equation}
P^{sp} = \int 2\gamma\left|\beta(t)\right|^2  dt= \gamma/\gamma_c\:,
\end{equation}
which is independent of $\kappa_s$. The efficiency of the routing is thus not increased by taking longer and longer pulses, as is the case for stimulated Raman adiabatic passage, but can be suppressed by increasing the cavity-enhanced atomic decay rate. The effect of intrinsic losses of the microtoroid cavity can be evaluated by omitting the fast-cavity approximation, and including the microtoroid cavity modes in the system dynamics. The probability of losing a photon is then calculated to be
\begin{equation}
P^{loss} = 4\kappa_i/\kappa_{ex}.
\end{equation}
This effect can be suppressed by increasing the cavity decay rate into the fiber.
Hence, by taking $\gamma_c \gg \gamma$ and $\kappa_{ex} \gg \kappa_i$, the routing mechanism is made robust against both loss mechanisms.
\\

\section{\label{sec:level5}Conclusion}

In this work we analyzed photon routing in cavity-QED, focusing on the task of separating two incoming indistinguishable photons to different ports. We have shown that using just one cavity-enhanced two-level system is insufficient for this task, since the bandwidth of the system and the timescale of its nonlinearity, namely the temporal duration at which the system will respond differently to two photons than to one photon, are inherently linked by the uncertainty principle. This inherent conflict limits the efficiency of photon routing to~$\sim64\%$. We have shown that even the use of time-energy entangled photons at the input pulse does not circumvent this conflict completely, although it does enable some increase in the efficiency of the process. Finally, we have shown how the use of a three-level system does enable ideal photon routing, free from the limitations of a two-level system. Specifically, we presented a scheme that uses an atom in the $\Lambda$-configuration with both transitions enhanced by a single-sided cavity, to create a photon routing mechanism in which one and only one photon from the incoming pulse is directed to one port, while the remaining photons are directed to the other port. This scheme is robust against variations in the pulse width and parasitic losses with realistic parameters, and provides a promising method for efficient routing of single photons.\\

\acknowledgments
This work was partially supported by the Israel Science Foundation, the Wolfson Foundation and the Crown Photonics Center. This research was made possible in part by the historic generosity of the Harold Perlman Family. SP acknowledges support from the Marsden Fund of the Royal Society of New Zealand.

\appendix
\section{}

The two-photon state of the cascaded system depicted in Fig.~\ref{fig:cascaded} can be decomposed into
\begin{equation}
\label{state}
\left|\psi(t)\right\rangle=\alpha(t)\left|2gg\right\rangle+\beta(t)\left|1eg\right\rangle
                          +\delta(t)\left|1ge\right\rangle+\eta(t)\left|0ee\right\rangle,
\end{equation}
where the second and third indices describe the atomic states of the single-sided subsystem and the photon router subsystem, respectively.
Using the Schr\"{o}dinger equation, the coefficients of Eq.~\eqref{state} are found to be
\begin{widetext}
\begin{eqnarray}
\alpha(t) &=& e^{-2\kappa_s t}\\
\beta(t) &=&  -2\sqrt{2}\frac{\sqrt{\gamma_{c1}\kappa_s}}{\gamma_{c1}-\kappa_s}\left[e^{-\kappa_s t} - e^{-\gamma_{c1} t}\right]e^{-\kappa_s t}\nonumber\\
\delta(t) &=& 2\sqrt{2}\frac{\gamma_{c1}+\kappa_s}{\gamma_{c1}-\kappa_s}\frac{\sqrt{\gamma_{c2}\kappa_s}}{2\gamma_{c2}-\kappa_s}e^{-2\kappa_s t}
-4\sqrt{2}\frac{\gamma_{c1}}{\gamma_{c1}-\kappa_s}\frac{\sqrt{\gamma_{c2}\kappa_s}}{2\gamma_{c2}-\gamma_{c1}}e^{-(\gamma_{c1}+\kappa_s)t}
+2\sqrt{2}\frac{2\gamma_{c2}+\gamma_{c1}}{2\gamma_{c2}-\gamma_{c1}}\frac{\sqrt{\gamma_{c2}\kappa_s}}{2\gamma_{c2}-\kappa_s}e^{-(2\gamma_{c2}+\kappa_s)t}
\nonumber\\
\eta(t) &=& 4\sqrt{2}\frac{\kappa_s\sqrt{\gamma_{c1}\gamma_{c2}}}{(\gamma_{c1}-\kappa_s)(2\gamma_{c2}-\kappa_s)}  \left\{
 \frac{2\gamma_{c2}-\gamma_{c1}-2\kappa_s}{2\gamma_{c2}+\gamma_{c1}-2\kappa_s}   \left[e^{-2\kappa_s t} - e^{-(2\gamma_{c2}+\gamma_{c1}) t}\right]  \right.\nonumber\\
  && \left. +\frac{3\gamma_{c1}-2\gamma_{c2}}{2\gamma_{c2}-\gamma_{c1}}                    \left[e^{-(\gamma_{c1}+\kappa_s) t} - e^{-(2\gamma_{c2}+\gamma_{c1})t}\right]
 -\frac{2\gamma_{c2}+\gamma_{c1}}{2\gamma_{c2}-\gamma_{c1}}                      \left[e^{-(2\gamma_{c2}+\kappa_s) t} - e^{-(2\gamma_{c2}+\gamma_{c1})t}\right] \right\}\nonumber.
\end{eqnarray}
\end{widetext}
After a photon detection, the system can collapse into three different single-excitation states.
If the photon collapses to the feeder cavity, the state evolves according to
\begin{equation}
\left|\psi_1(t)\right\rangle=a_1\left|1gg\right\rangle+b_1\left|0eg\right\rangle+c_1\left|0ge\right\rangle,
\end{equation}
where

\begin{eqnarray}
a_1(t) &=& e^{-\kappa_s t}\\
b_1(t) &=& -2\frac{\sqrt{\gamma_{c1}\kappa_s}}{\gamma_{c1}-\kappa_s}\left(e^{-\kappa_s t}-e^{-\gamma_{c1} t}\right)\nonumber\\
c_1(t) &=& 2\frac{\sqrt{\gamma_{c2}\kappa_s}}{2\gamma_{c2}-\kappa_s}\left\{
\frac{\gamma_{c1}+\kappa_s}{\gamma_{c1}-\kappa_s}e^{-\kappa_s t}
-2\frac{\gamma_{c1}}{\gamma_{c1}-\kappa_s}e^{-\gamma_{c1}t}   \right.\nonumber\\
&& \left.+\frac{2\gamma_{c2}+\gamma_{c1}}{2\gamma_{c2}-\gamma_{c1}}e^{-2\gamma_{c2}t}\right\}\nonumber.
\end{eqnarray}

If the system collapses to a state in which the first atom is excited, we have
\begin{equation}
\left|\psi_2(t)\right\rangle=b_2\left|0eg\right\rangle+c_2\left|0ge\right\rangle,
\end{equation}
where
\begin{eqnarray}
b_2(t) &=& e^{-\gamma_{c1} t}\nonumber\\
c_2(t) &=& -2\frac{\sqrt{\gamma_{c1}\gamma_{c2}}}{2\gamma_{c2}-\gamma_{c1}}\left(e^{-\gamma_{c1} t}-e^{-2\gamma_{c2} t}\right).
\end{eqnarray}
And finally, if the state collapses to a state in which the second atom is excited, we have $\left|\psi_3(t)\right\rangle=c_3(t)\left|0ge\right\rangle$, with
\begin{equation}
c_3(t) = e^{-2\gamma_{c2} t}.
\end{equation}
Using Eq.~\eqref{operators}, the correlation functions are given by
\begin{widetext}
\begin{eqnarray}
\Gamma^{tr}(t,\tau) &=& 4\left|
       \sqrt{\gamma_{c2}}\eta(t)\left[\sqrt{\gamma_{c1}}b_2(\tau)+\sqrt{\gamma_{c2}}c_2(\tau)\right] +
       \sqrt{\gamma_{c2}}\delta(t)\left[\sqrt{\kappa_s}a_1(\tau)+\sqrt{\gamma_{c1}}b_1(\tau)+\sqrt{\gamma_{c2}}c_1(\tau)\right]
                               \right|^2  \\
\Gamma^{rt}(t,\tau) &=& 4\left|
      \sqrt{\gamma_{c2}}\left[\sqrt{2\kappa_s}\alpha(t)+\sqrt{\gamma_{c1}}\beta(t)+\sqrt{\gamma_{c2}}\delta(t)\right]c_1(\tau)\right.\nonumber\\
      &&\left.+\sqrt{\gamma_{c2}}\left[\sqrt{\gamma_{c2}}\eta(t)+\sqrt{\kappa_s}\beta(t)\right]c_2(\tau)
        + \sqrt{\gamma_{c2}}\left[\sqrt{\gamma_{c2}}\eta(t)+\sqrt{\kappa_s}\delta(t)\right]c_3(\tau)
                               \right|^2   \nonumber\\
\Gamma^{rr}(t,\tau) &=& 4\left|\gamma_{c2}
        \left(  \eta(t)c_2(\tau)+ \delta(t)c_1(\tau) \right)
                             \right|^2     \nonumber\\
\Gamma^{tt}(t,\tau) &=& 4\left|
        \left[\sqrt{2\kappa_s}\alpha(t)+\sqrt{\gamma_{c1}}\beta(t)+\sqrt{\gamma_{c2}}\delta(t)\right]\left[\sqrt{\kappa_s}a_1(\tau)+\sqrt{\gamma_{c1}}b_1(\tau)+\sqrt{\gamma_{c2}}c_1(\tau)\right]\right. \nonumber\\
        &&\left.+\left[\sqrt{\gamma_{c2}}\eta(t)+\sqrt{\kappa_s}\beta(t)\right]\left[\sqrt{\gamma_{c1}}b_2(\tau)+\sqrt{\gamma_{c2}}c_2(\tau)\right]
         +\sqrt{\gamma_{c2}}\left[\sqrt{\gamma_{c2}}\eta(t)+\sqrt{\kappa_s}\delta(t)\right]c_3(\tau)
                              \right|^2.\nonumber
\end{eqnarray}
\end{widetext}
Integrating these expressions yields the results plotted in Fig.~\ref{fig:gc1gc2}.

\end{document}